\documentclass{article}
\usepackage{PRIMEarxiv}

\usepackage[utf8]{inputenc} 
\usepackage[T1]{fontenc}    
\usepackage{hyperref}      
\usepackage{url}           
\usepackage{booktabs}       
\usepackage{amsfonts}       
\usepackage{nicefrac}      
\usepackage{microtype}      
\usepackage{lipsum}
\usepackage{fancyhdr}       
\usepackage{graphicx}      
\graphicspath{{media/}}   
\usepackage{xcolor}

\title{Ethical Framework for Responsible Foundational Models in Medical Imaging} 

\usepackage[utf8]{inputenc}
\usepackage{authblk}

\title{Ethical Framework for Responsible Foundational Models in Medical Imaging}

\author[1]{Debesh Jha}
\author[1]{Gorkem Durak}
\author[1]{Abhijit Das}
\author[1]{Jasmer Sanjotra}
\author[1]{Onkar Susladkar}
\author[1]{Suramyaa Sarkar}
\author[2]{Ashish Rauniyar}
\author[1]{Nikhil Kumar Tomar}
\author[1]{Linkai Peng}
\author[1]{Sirui Li}
\author[1]{Koushik Biswas}
\author[1]{Ertugrul Aktas}
\author[1]{Elif Keles}
\author[1]{Matthew Antalek}
\author[1]{Zheyuan Zhang}
\author[1]{Bin Wang}
\author[1,3]{Xin Zhu}
\author[1]{Hongyi Pan}
\author[1]{Deniz Seyithanoglu}
\author[1]{Alpay Medetalibeyoglu}
\author[1]{Vanshali Sharma}
\author[1]{Vedat Cicek}
\author[1]{Amir A. Rahsepar}
\author[1,9]{Rutger Hendrix}
\author[3]{A. Enis Cetin}
\author[4]{Bulent Aydogan}
\author[5]{Mohamed Abazeed}
\author[1]{Frank H. Miller}
\author[1,6]{Rajesh N. Keswani}
\author[1]{Hatice Savas}
\author[7]{Sachin Jambawalikar}
\author[8]{Daniela P. Ladner}
\author[1]{Amir A. Borhani}
\author[1,9]{Concetto Spampinato}
\author[1,10]{Michael B. Wallace}
\author[1]{Ulas Bagci\thanks{Corresponding author: ulas.bagci@northwestern.edu}}

\affil[1]{Machine and Hybrid Intelligence Lab, Department of Radiology, Northwestern University, Chicago, IL, USA}
\affil[2]{Sustainable Communication Technologies, SINTEF Digital, Trondheim, Norway}
\affil[3]{Department of Electrical and Computer Engineering, University of Illinois at Chicago, Chicago, IL, USA}
\affil[4]{Department of Radiation Oncology, University of Chicago, Chicago, IL, USA}
\affil[5]{Department of Radiation Oncology, Northwestern University, Chicago, IL, USA}
\affil[6]{Department of Gastroenterology and Hepatology, Northwestern University, Chicago, IL, USA}
\affil[7]{Department of Radiology, Columbia University, NYC, NY, USA}
\affil[8]{Northwestern University Transplant Outcomes Research Collaborative (NUTORC), Comprehensive Transplant Center, Feinberg School of Medicine, Northwestern University, Chicago, IL, USA}
\affil[9]{Department of Electrical and Computer Engineering, University of Catania, Catania, Italy}
\affil[10]{Department of Gastroenterology and Hematology, Mayo Clinic Florida, Jacksonville, FL, USA}

\date{}

\begin{document}
\maketitle

\begin{abstract}
The emergence of foundational models represents a paradigm shift in medical imaging, offering extraordinary capabilities in disease detection, diagnosis, and treatment planning. These large-scale artificial intelligence systems, trained on extensive multimodal and multi-center datasets, demonstrate remarkable versatility across diverse medical applications. However, their integration into clinical practice presents complex ethical challenges that extend beyond technical performance metrics. This study examines the critical ethical considerations at the intersection of healthcare and artificial intelligence. Patient data privacy remains a fundamental concern, particularly given these models' requirement for extensive training data and their potential to inadvertently memorize sensitive information.  Algorithmic bias poses a significant challenge in healthcare, as historical disparities in medical data collection may perpetuate or exacerbate existing healthcare inequities across demographic groups. The complexity of foundational models presents significant challenges regarding transparency and explainability in medical decision-making.  We propose a comprehensive ethical framework that addresses these challenges while promoting responsible innovation. This framework emphasizes robust privacy safeguards, systematic bias detection and mitigation strategies, and mechanisms for maintaining meaningful human oversight. By establishing clear guidelines for development and deployment, we aim to harness the transformative potential of foundational models while preserving the fundamental principles of medical ethics and patient-centered care.
\end{abstract}


\section{Introduction}
\label{sec:introduction}
{Recent advancements in artificial intelligence (AI) have been catalyzed by the emergence of \textbf{foundational models (FMs)}~\cite{medetalibeyoglu2025foundational}—large-scale architectures capable of generalizing across diverse applications with significantly reduced data requirements compared to traditional deep learning paradigms~\cite{brown2020language,statnews2022epicsepsis}.} 
These models, which leverage massive parameter spaces and extensive training datasets, have achieved remarkable performance even when using only a tenth of the conventional data volume~\cite{rasmy2021med}. 
{This transformative progress is largely driven by two key innovations: (1) the convergence of high-performance computing and scalable parallel architectures and (2) the adoption of self-supervised learning strategies, particularly those based on the transformer architecture \cite{vaswani2017attention}.}

{\subsection*{The Evolution of Foundational Models in Medical Imaging}}
The theoretical underpinnings of FMs rest on two key machine learning paradigms: transfer learning~\cite{thrun1998lifelong} and unsupervised learning~\cite{bommasani2021opportunities}. While traditional medical imaging has relied heavily on vision-specific architectures such as convolutional neural networks (CNNs) and vision transformers, these approaches face significant limitations~\cite{chen2021transunet,diakogiannis2020resunet}. The conventional fully-supervised learning paradigm demands substantial annotated datasets, making it resource-intensive and time-consuming. Furthermore, these models typically specialize in single tasks, such as segmentation or classification, and operate within a single modality.

This single-modality constraint presents a fundamental mismatch with real-world healthcare workflows, where clinicians routinely integrate multiple information sources including clinical notes, diagnostic reports, and various investigative findings to make informed decisions. FMs for computer-aided diagnosis (CAD) represent a strategic shift toward addressing these limitations while maintaining crucial considerations of patient privacy, model transparency, and ethical implementation. The evolution from traditional deep learning approaches to FMs mirrors the complexity of actual clinical decision-making, where the synthesis of diverse information sources drives diagnostic accuracy and treatment planning~\cite{duggan2021improving,stanford_hai_foundation_models_2024}.
{This transition is particularly consequential in medical imaging, where diagnostic accuracy is contingent on integrating heterogeneous information. For example, radiologists rely on multimodal inputs—imaging scans, clinical histories, and laboratory results—to refine differential diagnoses. FMs hold the potential to revolutionize this process by enhancing diagnostic precision, automating complex tasks, and personalizing treatment strategies at an unprecedented scale.}

\begin{figure*}[!t]
    \centering
    \includegraphics[width=1\linewidth]{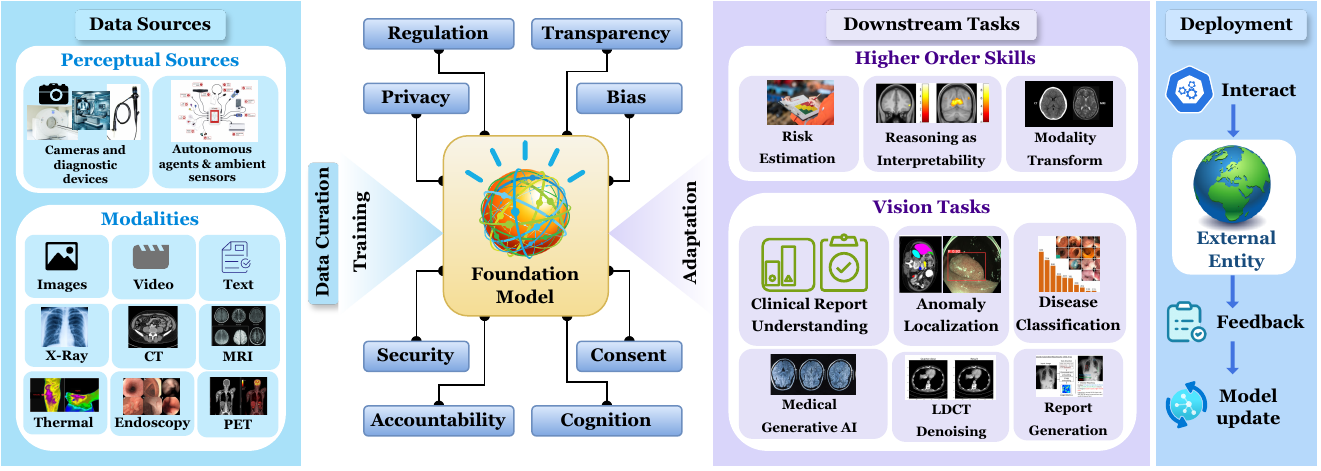}
    \caption{Comprehensive workflow of a foundational model in medical imaging: from multi-modal data acquisition and curation to deployment and downstream tasks for higher order skills and vision tasks.}
    \label{fig:diagram}
\end{figure*}

{\subsection*{Ethical and Practical Challenges}}
Despite the remarkable achievements of FMs and large vision models (LVMs) in medical applications~\cite{lei2023medlsam,ma2024segment}, their widespread adoption raises significant ethical and societal concerns that demand careful consideration. The substantial data requirements for training these models present complex challenges regarding patient privacy and data confidentiality. Medical datasets contain highly sensitive information, including detailed health histories and genetic data, necessitating robust protection mechanisms beyond traditional security measures. A more nuanced challenge emerges from inherent biases within training datasets. These biases can manifest in various forms, potentially leading to discriminatory outcomes based on demographic factors such as race, gender, and socioeconomic status. Such biases not only compromise diagnostic accuracy but also risk perpetuating existing healthcare disparities when deployed in clinical settings. The accountability for these biased outcomes becomes particularly complex given the multiple stakeholders involved in developing and deploying medical FMs.

The generative capabilities of modern FMs introduce additional layers of ethical complexity, particularly regarding potential misuse and legal liability. The inherent opacity of these sophisticated models, often characterized as ``blackbox'', {necessitates advanced explainable AI techniques to establish trust among healthcare providers and patients alike. This transparency is crucial for clinical adoption and regulatory compliance. Hence, FMs in medical imaging face several interconnected challenges, summarized briefly as follows:}\\

\noindent\textbf{\textit{(i) Data scarcity.}} A fundamental constraint lies in the scarcity of high-quality annotated medical images, which limits the training capabilities of these sophisticated models.\\
\textbf{\textit{(ii) Variation.}} This challenge is compounded by the inherent complexity of medical imaging data, where high-resolution volumetric scans display significant anatomical variations between individuals, making it difficult to develop models that generalize effectively across diverse patient populations.\\
\textbf{\textit{(iii) Heterogeneous data.}} The heterogeneous nature of medical imaging data presents another layer of complexity. Healthcare facilities utilize various imaging devices and follow different protocols, resulting in a diverse array of data formats and characteristics. This variability in imaging modalities and acquisition parameters creates substantial challenges for developing unified models that can process and interpret such diverse inputs effectively.\\
\textbf{\textit{(iv) Computational cost.}} Scalability emerges as a critical operational challenge in implementing medical FMs. These sophisticated models demand substantial computational resources, leading to extended processing times and increased operational costs. This resource-intensive nature can potentially limit their practical deployment in clinical settings where rapid analysis and cost-effectiveness are crucial considerations.\\
\textbf{\textit{(v) Ethics and reliability.}} Beyond these technical challenges, ethical considerations and reliability concerns pose significant hurdles. The handling of sensitive patient data necessitates robust privacy and security measures while ensuring data integrity remains paramount. The reliability of FM outputs faces particular scrutiny in medical contexts, where the stakes are exceptionally high. \\
\textbf{\textit{(vi) Susceptibility.}} Moreover, these models' vulnerability to adversarial attacks raises serious concerns~\cite{maus2023black}, given that medical decisions can have profound implications for patient outcomes.\\
These challenges span both domain-specific and general considerations~\cite{azad2023foundational}. Table~\ref{tab:challenges} presents a real-world example and the corresponding solution for each challenge.

{\subsection*{A Framework for Ethical AI in Medicine}
To address these challenges, we propose a comprehensive ethical framework integrating federated learning, bias mitigation techniques, and explainability modules. This framework emphasizes:}
{
\begin{enumerate}
\item \textbf{Ethical AI Development:}  We present an ethical framework that guides the responsible development and implementation of FMs in medicine. We propose to implement privacy-preserving methodologies such as homomorphic encryption and decentralized learning to protect patient confidentiality.
\item \textbf{Fairness \& Equity:} Establishing robust bias detection and mitigation strategies to prevent discriminatory outcomes.
\item \textbf{Transparency \& Clinical Trust:} Leveraging interpretable AI mechanisms and clinician-AI collaboration to foster adoption and regulatory compliance.
\end{enumerate}}
{This work aims to set the foundation for responsible AI integration in medicine, ensuring that FMs enhance clinical decision-making without compromising ethical integrity or patient safety. The innovation of this paper lies in its comprehensive ethical framework for medical FMs, integrating privacy-preserving techniques (e.g., federated learning, homomorphic encryption), fairness-aware training, and explainable AI to address critical challenges in medical AI deployment. Unlike conventional deep learning models that rely on single-task, single-modality architectures, this work presents a framework with a multi-modal, multi-task paradigm that aligns with real-world clinical decision-making. Additionally, we propose a systematic bias auditing and regulatory compliance strategy, ensuring that FMs promote equitable, transparent, and trustworthy AI-driven healthcare.}

\begin{table}[htbp]
\centering
\caption{Challenges, Examples, and Solutions in Medical Imaging.}
\begin{small}
\begin{tabular}{p{2.1cm}|p{5cm}|p{9.3cm}}
\toprule
\textbf{Challenge} & \textbf{Real-World Example} & \textbf{Real-World Solution} \\ \midrule
Data scarcity & A rare disease imaging dataset has only a few dozen annotated examples, making it difficult to train a robust AI model for diagnosis. & Utilize transfer learning by leveraging pre-trained models on large general medical imaging datasets and fine-tune them for rare diseases. Use data augmentation techniques (e.g., rotation, flipping, scaling) to artificially increase the diversity of the dataset. \\ \hline
Variation & Chest X-rays from patients of different ethnicities show significant differences in anatomical features and disease manifestations, leading to inconsistent model performance. & Train models on diverse and representative datasets that include data from multiple demographics. Implement domain adaptation techniques to improve the model's generalization across varied patient populations. Regular validation on diverse test sets is essential. \\ \hline
Heterogeneous data & MRI scans from different hospitals vary due to different imaging protocols, machine types, and acquisition settings, causing challenges in creating a standardized analysis model. & Develop and apply normalization and harmonization techniques to preprocess data to a common format and quality. Use federated learning to train models on decentralized data while maintaining patient privacy and improving model robustness across heterogeneous data sources. \\ \hline
Computational cost & Running a large AI model to analyze MRI scans is slow on high-end hardware, delaying critical diagnoses in emergency scenarios. & Optimize models using techniques such as model pruning and quantization to reduce size and computation. Incorporate edge computing for real-time analysis where possible and leverage cloud-based platforms with scalable resources for handling large-scale computations. \\ \hline
Ethics and reliability & A misdiagnosis by an AI system in detecting a malignant tumor could lead to incorrect treatment, raising ethical and trust issues among clinicians and patients. & Implement rigorous validation and explainability mechanisms to ensure transparency and reliability. Incorporate human-in-the-loop systems where clinicians review and validate AI predictions. Establish robust patient consent protocols and maintain high standards of data encryption and privacy measures to ensure compliance with healthcare regulations. \\ \hline
Susceptibility & An adversarial attack alters a medical image subtly, causing the AI model to misclassify a benign condition as malignant, leading to unnecessary surgeries. & Enhance model security through adversarial training, where the model is exposed to and learns from adversarial examples during training. Monitor model outputs for anomalies and use robust verification systems to flag unexpected predictions for human review. Regularly update models to defend against emerging adversarial techniques. \\ \bottomrule
\end{tabular}\end{small}
\label{tab:challenges}
\end{table}

In the following sections, we provide a detailed examination of these challenges and their implications for the development and deployment of medical imaging FMs. This analysis serves as a foundation for understanding the complex landscape of AI implementation in healthcare.

\section{Method}
The societal implications of FMs in healthcare extend beyond individual applications, encompassing broader ecosystem-wide effects that scale with model deployment. As illustrated in Figure~\ref{fig:diagram}, the ethical considerations surrounding large-scale FM adoption in medical settings present both challenges and opportunities for systematic improvement. These ethical dimensions can be systematically evaluated and optimized through quantifiable metrics that promote transparency, maintain data integrity, and ensure equitable outcomes across diverse patient populations.

Our comprehensive analysis and subsequent proposals establish a robust framework for developing and implementing ethically sound FMs in biomedical artificial intelligence. This framework addresses not only the technical aspects of model development but also the broader societal responsibilities inherent in deploying AI systems in healthcare. By focusing on measurable ethical criteria and clear governance structures, we aim to create a sustainable and responsible FM ecosystem that serves the healthcare community while protecting patient interests. This approach represents a critical step toward harmonizing technological advancement with ethical imperatives in medical AI, setting a foundation for future developments that prioritize both innovation and responsibility. The following sections detail our analysis and recommendations for achieving this balance.

\subsection{Glass box FMs: Towards transparency}

The growing emphasis on \textit{glass-box models} in healthcare represents a crucial shift toward interpretable artificial intelligence, addressing major requirements for trust and transparency in medical decision-making~\cite{franzoni2023black}. These models provide essential insights into their decision-making processes, making them particularly valuable in clinical settings where understanding the reasoning behind AI recommendations is paramount. Healthcare professionals' confidence in AI systems fundamentally depends on their ability to comprehend the underlying decision mechanisms. This transparency enables clinicians to effectively integrate AI assistance into their practice while maintaining their professional judgment and accountability. Similarly, patient acceptance of AI-driven healthcare recommendations significantly increases when the decision-making process is transparent and comprehensible, fostering a more trusting relationship between patients, healthcare providers, and AI systems.

Several sophisticated tools and techniques have emerged to enhance the interpretability of foundational models in commercial medical applications. These include Gradient-weighted Class Activation Mapping (CAM) methods~\cite{selvaraju2017grad,chattopadhay2018grad}, which visualize regions of interest in medical images that influence model decisions. Principle component analysis offers a gradient-independent approach to understanding data patterns~\cite{jolliffe2016principal}, while SHAP (SHapley Additive exPlanations)~\cite{lundberg2017unified} and LIME (Local Interpretable Model-agnostic Explanations)~\cite{ribeiro2016should} provide detailed insights into model predictions. These visual reasoning techniques collectively enable a deeper understanding of how FMs process and analyze medical data, making their decisions more transparent and trustworthy for both healthcare providers and patients.

A clinical scenario where the model trained on a biased model due to an imbalanced training dataset. For example, consider a model trained predominantly on male patients with hypertrophic cardiomyopathy (HCM). When deployed in the real world, the model may fail to detect HCM in female patients due to underlying gender-based biases in the training data. By incorporating interpretable AI, clinicians can identify and understand these biases. For instance, interpretable AI might reveal that the model underweights key diagnostic features in female patients. This insight allows physicians to adjust their clinical decisions and provides feedback to retrain the model with more diverse and representative data, thereby improving future diagnostic accuracy and reducing gender-related disparities.

\subsection{Federated learning: Ensuring privacy}
\begin{figure} [!t]
    \centering
    \includegraphics[width=0.8\linewidth]{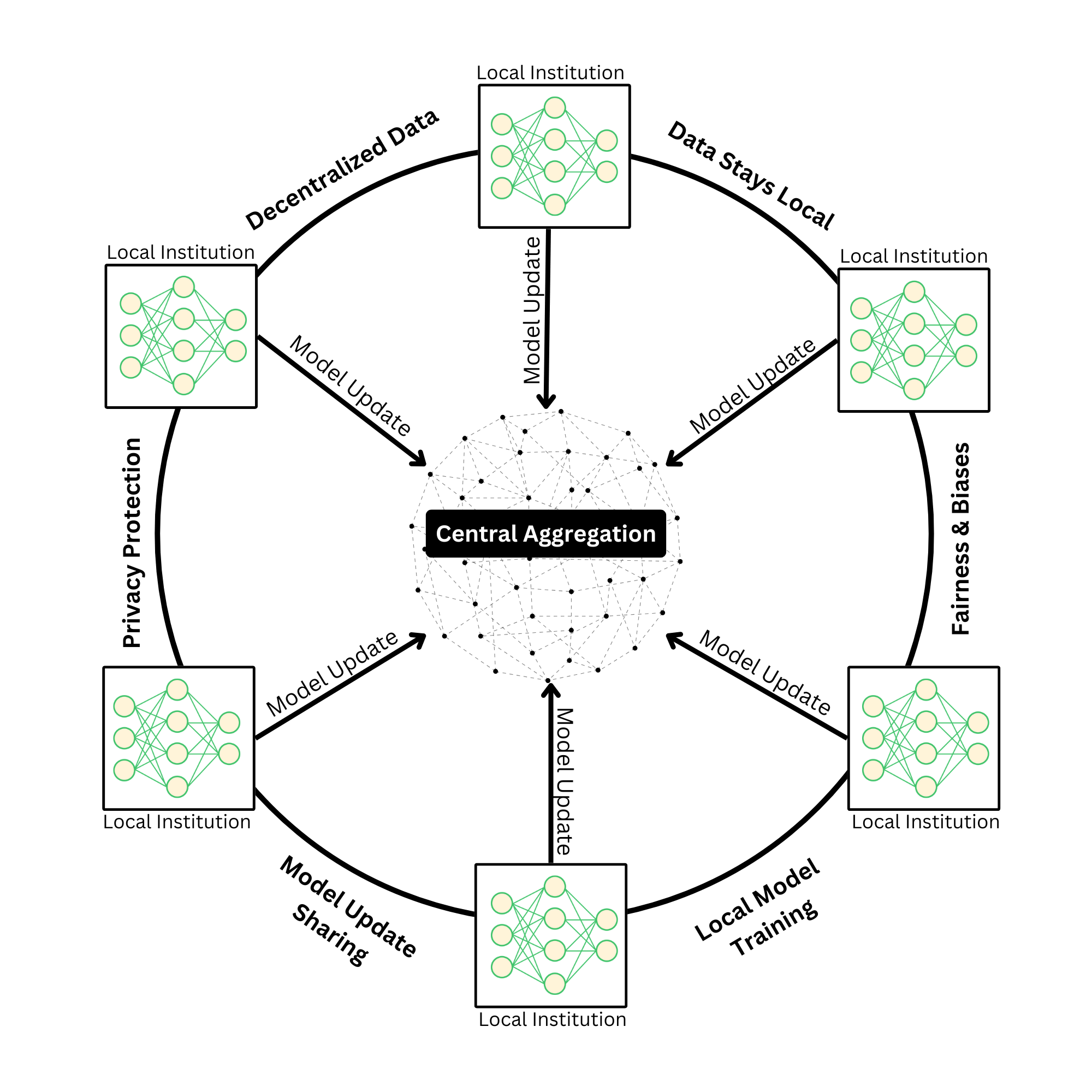}
    \caption{{Illustration of the \textbf{federated learning} paradigm, in which multiple institutions collaboratively train deep learning models in a decentralized framework without exchanging raw patient data. Each institution independently updates its local model using private datasets and transmits only model parameters to a central aggregation server. The server securely integrates these updates to refine a global model, which is then redistributed to participating sites. This iterative process preserves data privacy, maintains data locality, and mitigates risks related to bias and fairness while ensuring robust model generalization across diverse clinical settings.}} 
    \label{fig:fl}
    \vspace{-5mm}
\end{figure}

The exceptional performance of FMs in medical applications heavily depends on access to extensive, high-quality training data. However, the medical field faces a critical challenge in data availability, particularly given the sensitive nature of patient information and the time-intensive process of curating private medical datasets. This constraint has led to the emergence of federated learning as a transformative solution for medical AI development. Federated learning represents a paradigm shift in how medical FMs can be trained while preserving patient privacy. This approach enables the development of robust models by leveraging distributed data sources across multiple healthcare institutions without requiring centralized data storage (Figure~\ref{fig:fl}). The key innovation lies in its ability to keep sensitive patient data securely within its original location while allowing the model to learn from multiple sources simultaneously. This distributed architecture addresses not only privacy concerns but also regulatory compliance requirements in healthcare.

Moreover, federated learning provides an elegant solution to the Non-IID (Non-Independent and Identically Distributed)~\cite{zhao2018federated} challenge that frequently occurs in medical datasets. By implementing granular controls over data sharing and model updates, healthcare institutions can maintain oversight of their contributions while benefiting from collaborative learning. This approach facilitates the development of more inclusive and representative models by incorporating diverse patient populations across different healthcare settings. The resulting federated FMs demonstrate enhanced fairness and reduced bias, as they can learn from a broader spectrum of medical data while respecting privacy boundaries and institutional protocols.

\subsection{LLMs: Facilitating regulatory compliance} 
Large Language Models (LLMs) are revolutionizing computer-aided diagnosis (CAD) systems by bridging the gap between visual analysis and clinical documentation. Models like LLaMa and Komodo-7b~\cite{owen2024komodo} demonstrate remarkable capabilities in transforming unstructured medical information into comprehensive, standardized formats. This transformation extends beyond simple text generation~\cite{shi2023llm}, encompassing crucial healthcare applications including the creation of detailed Electronic Health Records (EHRs), clinical trial analysis, drug discovery processes, biomarker identification, and the enhancement of Clinical Decision Support Systems (CDSS)~\cite{mishuris2014using}.

The integration of LLMs into healthcare workflows addresses critical regulatory compliance requirements while improving documentation efficiency. These models excel at generating structured medical records that adhere to stringent privacy and security regulations, significantly reducing the risk of non-compliance penalties. The implementation of sophisticated privacy-preserving techniques, such as differential privacy, adds an essential layer of security by introducing controlled noise into training data, thereby protecting patient confidentiality while maintaining data utility.

The ongoing clinical trials of LLM applications in healthcare settings serve a dual purpose: validating their effectiveness in real-world scenarios and ensuring compliance with regulatory frameworks, particularly the Health Insurance Portability and Accountability Act (HIPAA). This rigorous evaluation process helps establish LLMs as reliable tools that can enhance healthcare delivery while maintaining the highest standards of patient privacy and data security. The successful integration of these models demonstrates how advanced AI technologies can support healthcare professionals in delivering more efficient and compliant care.

\subsection{Generative AI: Generalization with privacy}
Generative models~\cite{goodfellow2014generative} have emerged as a powerful solution to several fundamental challenges in medical AI, particularly addressing the critical issue of data scarcity in training foundational models (FMs). These models excel at creating synthetic medical data that closely mirrors real patient information, effectively expanding training datasets while circumventing privacy and consent concerns inherent in using actual patient data. By generating diverse synthetic samples that represent various demographic and clinical characteristics, these models help establish more balanced and representative training datasets.

Variational autoencoders (VAEs)~\cite{kingma2013auto} represent a particularly sophisticated application of generative modeling in healthcare. Their ability to predict missing values and generate synthetic patient trajectories enhances the robustness of FMs by providing more complete and diverse training data\cite{esmaeili2023generative}. This capability proves especially valuable in medical settings where incomplete or missing data often poses significant challenges to model development and deployment.

Recent advances in self-supervised learning have further enhanced the potential of generative approaches. Notable work by Ghesu and colleagues demonstrated the effectiveness of combining contrastive learning with online feature clustering for dense feature learning in FMs~\cite{ghesu2201self}. Their hybrid approach, building upon earlier self-supervised techniques, achieves robust feature representations by mapping them to cluster prototypes through both supervised and self-supervised learning mechanisms~\cite{caron2020unsupervised}.

The integration of generative techniques with FMs has yielded remarkable results~\cite{susladkar2023tpfnet,deshmukh2024textual}, as exemplified by models like MedSAM~\cite{ma2024segment}, which demonstrates superior performance through generative AI-based encoding-decoding architectures. This success extends to applications in generative image modeling, where synthetic data is used for pretraining and inference on real-world medical data, leading to optimized FM performance. These advances not only improve model accuracy but also incorporate crucial ethical considerations by emphasizing privacy-preserving data generation methods and bias reduction strategies.

\subsection{Fairness, biases, and risks with generative models}
The transformative potential of generative AI in healthcare is accompanied by significant ethical challenges that demand careful consideration. These models can inadvertently amplify existing social biases across multiple dimensions including race, gender, and socioeconomic status, potentially leading to discriminatory outcomes in medical decision-making~\cite{zhu2024could}. The sophisticated nature of these technologies raises particular concerns about their role in perpetuating or exacerbating existing healthcare disparities. The risk extends beyond bias amplification to include more direct threats to public trust and safety. The capability of generative AI to create convincing deepfakes and propagate medical misinformation presents serious challenges to healthcare communication and patient trust. These issues are particularly concerning in medical contexts, where accurate information is crucial for patient care and public health decisions. The potential for societal harm increases when these technologies trigger public hostility or erode trust in healthcare institutions~\cite{masood2023deepfakes}.

Addressing these challenges requires a comprehensive approach that prioritizes ethical considerations over purely technological advancement. Organizations developing medical AI systems must shift their focus from maximizing model performance to actively minimizing bias and potential harm. This paradigm shift emphasizes the importance of building trustworthy systems that serve all populations equitably, rather than pursuing technological capabilities at the expense of ethical considerations. The path forward requires early intervention in AI education and development, embedding responsible usage principles at fundamental stages of both technical training and clinical implementation. This approach must also address the broader socioeconomic implications of AI deployment in healthcare, particularly the risk of creating or widening digital divides that favor well-resourced healthcare systems while potentially disadvantaging others. Success in this endeavor demands active collaboration among healthcare providers, AI developers, policymakers, and patient advocates to ensure that generative AI advances medical care while upholding ethical principles and promoting equitable access.

\subsection{Methods for measuring fairness, bias, privacy, and diversity of generations}
The development of ethical generative AI systems in healthcare demands a rigorous approach to ensuring fairness and equity in model outcomes. A fundamental principle is that these systems should deliver consistent results for similar medical cases, independent of demographic factors such as race, gender, or socioeconomic status. This objective necessitates the implementation of sophisticated fairness metrics and systematic algorithmic audits to identify and address potential biases in both training data and model outputs.

Privacy protection in medical AI systems can be achieved through a multi-layered approach combining advanced techniques such as data anonymization with strategic noise injection and federated learning architectures~\cite{mehrabi2021survey}. These methods effectively minimize the risk of data breaches while maintaining model performance. The evaluation of model fairness employs quantitative measures such as the Gini coefficient and Shannon diversity index~\cite{ojha2022towards}, which provide objective metrics for assessing output diversity and detecting potential biases (Table~\ref{tab:metrics}). Higher diversity scores typically indicate more inclusive and less homogeneous model behavior across different demographic groups.

The integration of these evaluation techniques throughout the entire development life-cycle ensures continuous monitoring of fairness, bias, and diversity metrics~\cite{bommasani2021opportunities}. This systematic approach is essential for maintaining consistent performance across all patient populations. Achieving truly inclusive AI systems requires deliberate efforts to incorporate diverse representation in both training data and development teams, thereby preventing performance disparities that could disadvantage specific patient groups.

The challenge of addressing historical and societal biases in medical data requires a combination of technical solutions and social awareness~\cite{kuhlman2020no}. Through rigorous bias auditing and sophisticated debiasing techniques, developers can work to neutralize these embedded prejudices. Success in this endeavor requires meaningful collaboration between technologists, healthcare professionals, and social scientists, ensuring that medical AI systems serve all populations effectively and ethically.

\begin{table}[!t]
    \centering
    \caption{Methods for measuring fairness, bias, privacy, and diversity of generations.}
   \resizebox{\columnwidth}{!}{
    \begin{tabular}{l|c|c|c}
        \toprule
        \textbf{Method} & \textbf{Authors\&Year} & \textbf{Metric} & \textbf{Performance} \\  \midrule
        Fairness-Constrained&~\cite{hardt2016equality} & Equalized Odds, Demographic Parity & Fairness vs. accuracy trade-offs\\ \hline
        Fairness Auditing&~\cite{veale2017fairer} & Bias Detection & Continuous monitoring needed\\  \hline
        Gender Classification&~\cite{buolamwini2018gender} & Intersectional Accuracy & Varied accuracy; higher errors for darker-skinned females \\ \hline
        Federated Learning&~\cite{madras2019fairness} & Fairness, Privacy & Fairness with data privacy \\ \hline
        Private GANs&~\cite{seibold2020quantitative} & Privacy, Fidelity & Private data with good fidelity\\ \hline
        Multi-modal Foundation&~\cite{kairouz2021advances} & Gini Coefficient, Shannon Diversity & High diversity and balanced representations \\ \hline  
        Fair Representation&~\cite{barocas2023fairness} & Stat. Parity, Equalized Odds & Balanced fairness metrics \\ \bottomrule
    \end{tabular}}
    \label{tab:metrics}
\end{table}


\subsection{Copyright concerns}
The intersection of generative AI and copyright law presents complex challenges in medical imaging and healthcare applications. These AI systems' ability to generate content that may resemble existing work raises significant questions about intellectual property rights and fair use~\cite{lucchi2023chatgpt,chen2024generative}. The challenge becomes particularly nuanced in medical contexts, where the generated content could include diagnostic patterns, imaging techniques, or analytical methods that may be subject to existing patents or copyrights.

A balanced approach to addressing these concerns requires careful consideration of both innovation and protection. Healthcare AI developers must implement rigorous protocols to ensure their training methodologies respect intellectual property rights, including proper attribution of source materials and careful documentation of training data provenance. This challenge extends beyond simple compliance to fundamental questions about the ownership and rights associated with AI-generated medical insights and diagnostic tools~\cite{sag2023copyright,gans2024copyright}.

The evolving nature of AI technology necessitates new legal frameworks that can effectively address these emerging challenges while fostering innovation in healthcare. This requires sustained collaboration between multiple stakeholders: technologists who understand the technical capabilities and limitations of generative AI, legislators who can craft appropriate regulatory frameworks, and legal experts who can interpret and apply these frameworks in the context of existing intellectual property law~\cite{verma2023copyright}.

The path forward demands aggressive yet thoughtful action to establish clear guidelines for the ethical and legal implementation of generative AI in healthcare. These guidelines must balance the imperative for technological advancement in medical care with the protection of individual and institutional rights. Success in this endeavor requires a comprehensive approach that considers not only technical and legal aspects but also broader societal implications, ensuring that the development of medical AI serves the public good while respecting intellectual property rights.

\subsection{Governance and Collaboration}
The implementation of artificial intelligence in medical imaging demands a robust governance framework that places human oversight at its core, ensuring responsible and ethical decision-making throughout the AI life-cycle~\cite{onitiu2023limits,sharma2022artificial}. This framework must begin with design-based privacy principles that protect patient data from the earliest stages of development, embedding security and confidentiality into the fundamental architecture of AI systems~\cite{kumar2023understanding}.

The complexity of healthcare AI necessitates a multi-stakeholder approach to governance. By engaging diverse participants—including healthcare providers, patients, technologists, ethicists, and regulatory experts—the framework benefits from a rich tapestry of perspectives and experiences~\cite{gkontra2023challenges}. This inclusive approach helps identify potential challenges and opportunities that might be overlooked from a single viewpoint.

Safety in medical AI systems requires a comprehensive validation protocol that includes rigorous testing, continuous monitoring, and regular assessment of outcomes~\cite{zhang2023ethics}. The establishment of an Ethical Governance Council provides crucial oversight, ensuring that AI development and deployment align with established ethical principles and clinical standards~\cite{bankins2024multilevel}. This council serves as a guardian of patient interests while facilitating technological advancement.

Educational initiatives play a vital role in this framework by ensuring all stakeholders understand both the capabilities and limitations of AI systems. These awareness programs foster realistic expectations and promote responsible use of AI technologies in clinical settings~\cite{sharma2022artificial}. The framework also emphasizes continuous improvement, incorporating mechanisms to adapt AI systems as new data becomes available and medical knowledge advances~\cite{kumar2023understanding}.

\begin{figure} [!t]
    \centering
    \includegraphics[width=1\linewidth]{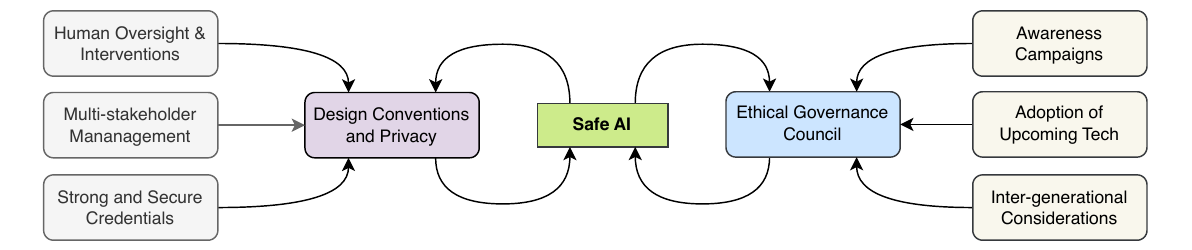}
    \caption{Framework of Ethical AI Governance Components.}
    \label{fig:governance}
\end{figure}

A particularly forward-thinking aspect of this governance structure is its consideration of intergenerational impacts. By addressing the needs of different age groups and anticipating future healthcare challenges, the framework ensures that AI development in medical imaging serves both current and future generations equitably~\cite{gkontra2023challenges}. As illustrated in Figure~\ref{fig:governance}, this comprehensive approach creates an ethical AI ecosystem that aligns technological innovation with societal values and healthcare needs~\cite{bankins2024multilevel}.

\subsection{Balance between scaling and societal impact for FMs} 
The advancement of artificial intelligence in healthcare requires careful navigation of interconnected practical and ethical challenges to ensure that technological innovation serves societal needs while minimizing potential harm. At the foundation of these challenges lies the critical issue of data quality and accessibility. The development of robust AI systems depends on access to diverse, representative datasets that capture the full spectrum of patient populations and medical conditions~\cite{sim2019focus}. However, this requirement must be balanced against stringent privacy requirements and ethical considerations in healthcare data management.

The technical challenge of developing scalable AI systems extends beyond pure computational capabilities to questions of seamless integration with existing healthcare infrastructure. These systems must operate efficiently within established clinical workflows while maintaining the highest standards of reliability and performance. This operational complexity is compounded by the imperative to maintain robust security measures that protect against data breaches and unauthorized access, particularly given the sensitive nature of medical information.

Bias mitigation represents one of the most pressing ethical challenges in medical AI development. The potential for AI systems to perpetuate or amplify existing healthcare disparities demands continuous innovation in fairness-ensuring techniques~\cite{baum2007looking}. This effort requires not only technical solutions but also deep understanding of how societal biases can manifest in healthcare data and decision-making processes. The development of transparent and accountable AI systems is crucial for building trust among healthcare providers and patients alike.

The broader societal implications of AI deployment in healthcare must be carefully considered and actively managed. This includes addressing concerns about potential job displacement in the medical sector, the responsible use of surveillance technologies, and the risk of exacerbating existing social inequalities in healthcare access. Success in navigating these challenges requires finding an optimal balance between technological advancement and societal acceptance, ensuring that AI development aligns with both clinical needs and public values.

\subsection{Security concerns and patient care} 
The emerging threat of ``jailbreaking'' in medical AI systems represents a critical vulnerability that extends beyond typical security concerns to potentially impact patient care directly. These unauthorized modifications of generative AI models can compromise the entire healthcare decision-making chain, introducing subtle yet dangerous alterations that may escape immediate detection~\cite{lapid2023open}. The implications of such tampering are particularly severe in medical imaging, where small alterations can lead to misdiagnosis or inappropriate treatment recommendations.

The risks associated with jailbreaking medical AI systems operate on multiple levels. At the technical level, these modifications can introduce systematic errors and biases that undermine the model's carefully calibrated performance. More critically, from a patient safety perspective, compromised systems may generate plausible-seeming but incorrect analyses, potentially leading to cascading errors in clinical decision-making. These technical vulnerabilities intersect with complex legal and regulatory requirements, potentially violating established healthcare standards and patient privacy protections~\cite{sun2024trustllm}.

The ethical implications of jailbreaking strike at the heart of fundamental medical principles. By compromising system integrity, these unauthorized modifications violate patient autonomy by potentially subjecting individuals to flawed medical decisions without their knowledge or consent. This breach of trust extends beyond individual patient relationships to potentially undermine broader public confidence in AI-driven healthcare solutions, threatening the advancement of beneficial medical AI applications.

Maintaining the integrity of medical AI systems requires a comprehensive defense strategy that prioritizes patient welfare above all other considerations. This necessitates collaboration between AI developers, healthcare providers, and regulatory bodies to establish robust security protocols and ethical guidelines. Only by maintaining an unwavering commitment to system integrity and patient safety can we preserve trust in AI-driven medical solutions and ensure their continued beneficial development~\cite{hannon2024robust}.

\subsection{Ethical and responsible use}
The development of ethical foundational models in healthcare requires a systematic approach to transparency and fairness that begins at the earliest stages of model development. Comprehensive documentation of data collection methodologies, preprocessing techniques, and model customization procedures creates a foundation of accountability and enables a thorough examination of potential biases. This documentation serves not only as a technical record but also as a crucial tool for identifying and addressing potential sources of bias before they can impact patient care.

Performance monitoring in healthcare AI must extend beyond traditional accuracy metrics to encompass fairness indicators across diverse patient populations. This requires sophisticated evaluation frameworks that can detect subtle performance variations across different demographic groups and clinical scenarios. The integration of advanced techniques such as data augmentation and algorithmic debiasing helps ensure that models maintain consistent performance across all patient populations, addressing potential disparities before they manifest in clinical practice~\cite{jin2023survey}.

Data protection in medical AI demands a multi-layered approach that combines technical solutions with rigorous governance protocols. The implementation of differential privacy techniques and federated learning architectures enables healthcare organizations to maintain high standards of data security while facilitating necessary model improvements. Regular security audits serve as critical checkpoints, identifying potential vulnerabilities and enabling proactive implementation of protective measures against emerging threats.

The concept of accountability in medical AI extends beyond technical performance to encompass broader responsibilities toward patient care and societal impact. This requires establishing clear chains of responsibility for AI-driven decisions and their consequences, creating channels for stakeholder feedback, and developing protocols for responsible model deployment. Success in this endeavor requires active engagement with external entities and a commitment to continuous improvement based on real-world performance and stakeholder input.

\section{Discussion}
{
\subsection*{Critical Analysis and Limitations}
Our framework for ethical FMs in medical imaging, while comprehensive, faces several critical challenges that warrant careful consideration. First, the inherent tension between model performance and interpretability remains largely unresolved. While we advocate for glass-box approaches, the increasing complexity of FMs often creates a trade-off between accuracy and explainability that cannot be easily reconciled with current technical solutions.} 

{The proposed federated learning approach, though promising for privacy preservation, introduces significant computational overhead and potential degradation in model performance. Healthcare institutions with varying computational resources and data quality may experience different levels of benefit from this distributed learning paradigm, potentially exacerbating existing healthcare disparities rather than mitigating them.}

{A critical limitation of our framework lies in its assumption of standardized data collection and annotation practices across healthcare institutions. The reality of medical data collection involves significant variability in protocols, equipment calibration, and annotation standards. This heterogeneity may undermine the effectiveness of our proposed bias detection and mitigation strategies.}

{
\subsection*{Practical Implementation Challenges and Regulations}
The implementation of our ethical framework faces several practical obstacles that require acknowledgment. The resource requirements for maintaining robust privacy measures and conducting comprehensive bias audits may be prohibitive for smaller healthcare facilities. This economic barrier could lead to a two-tiered system where only well-resourced institutions can fully implement ethical AI practices.
The proposed governance structure, while theoretically sound, may face resistance from various stakeholders. Clinicians may view additional oversight mechanisms as bureaucratic hurdles, while institutional administrators might resist the additional costs and complexity of implementing comprehensive ethical frameworks. These practical considerations could significantly impact the real-world adoption of our proposed solutions.}

{AI regulation is being shaped by a combination of international organizations and private tech giants, all of which are addressing the practical implementation challenges of ethical and responsible AI. UNESCO~\cite{unesco}, for example, focuses on global AI governance and ethical considerations, emphasizing the importance of human rights and transparency in AI deployment. Their initiatives highlight the difficulty of ensuring compliance across diverse regulatory environments. Meanwhile, the European Union (EU)~\cite{eu} is spearheading one of the most comprehensive AI regulatory efforts with its AI Act, which aims to classify and regulate AI systems based on risk levels. However, enforcement across EU member states poses logistical and legal challenges. At an intergovernmental level, the OECD~\cite{oecd} has established AI principles that emphasize fairness, transparency, and accountability, but translating these high-level guidelines into enforceable national policies remains a challenge. Private sector leaders are also taking steps toward AI regulation. Microsoft promotes "Responsible AI" frameworks, including bias mitigation and human oversight, but the challenge remains in integrating these ethical safeguards into rapidly evolving AI products. Similarly, Google’s AI principles~\cite{google} outline commitments to fairness and safety, but practical implementation is complicated by the need to balance innovation with regulation. Lastly, IBM’s focus on "Trustworthy AI" centers~\cite{ibm} on explainability and algorithmic fairness, yet the challenge lies in achieving industry-wide standardization while ensuring business viability. These varied approaches collectively aim to tackle the real-world obstacles of AI governance, but each faces difficulties in enforcement, standardization, and global applicability. The key challenge remains bridging the gap between regulatory ambition and practical implementation in AI development and deployment.}

{
\subsection*{Sociotechnical Considerations}
The broader societal implications of our framework deserve critical examination. The emphasis on technical solutions to ethical challenges may inadvertently overshadow the importance of human judgment and clinical expertise. There is a risk that over-reliance on automated systems, even those with built-in ethical safeguards, could gradually erode the human elements of healthcare delivery. Moreover, our approach to bias mitigation, while well-intentioned, may not adequately address the root causes of healthcare disparities. Technical solutions alone cannot resolve systemic inequities deeply embedded in healthcare systems and society at large. This limitation suggests the need for our framework to be integrated with broader systemic changes in healthcare delivery and medical education.
}

{
\subsection*{Future Research Directions and Open Questions}
Several critical questions remain unanswered and require further investigation:
\begin{enumerate}
    \item Scalability vs. Ethics: How can we balance the computational demands of ethical AI practices with the need for rapid clinical deployment?
    \item Governance Evolution: How should ethical frameworks adapt to emerging AI capabilities and evolving societal values?
    \item Cultural Considerations: How can our framework be adapted to different healthcare systems and cultural contexts while maintaining its ethical principles?
    \item Long-term Impact: What are the potential unintended consequences of widespread adoption of AI-driven medical imaging systems on healthcare profession dynamics?
\end{enumerate}
These questions highlight the need for ongoing critical evaluation and refinement of our framework.
}

\section{Conclusion}
Foundational models represent a pivotal advancement in medical imaging, promising to revolutionize diagnostic precision, treatment planning, and personalized medicine. Their potential to transform healthcare delivery extends beyond mere technical improvements, offering new possibilities for personalized medicine and enhanced clinical decision-making. However, this technological promise must be carefully balanced against the complex ethical challenges that emerge from their deployment in clinical settings. Our analysis reveals the multifaceted nature of these challenges, encompassing critical concerns about patient data privacy, algorithmic bias, model transparency, and professional accountability. The framework we propose addresses these challenges through a systematic approach that integrates technical solutions with ethical principles. By combining advanced privacy-preserving techniques, bias mitigation strategies, and robust accountability measures, we establish a foundation for responsible AI development in healthcare. The successful implementation of foundational models in medical practice demands unprecedented collaboration across disciplines. This includes not only technical experts and healthcare professionals but also ethicists, legal scholars, and patient advocates. Such diverse participation ensures that these powerful tools evolve in ways that respect patient rights, promote equitable care, and maintain the highest standards of medical ethics. The responsible development of medical AI requires constant vigilance and adaptation to emerging challenges. As these technologies continue to evolve, our ethical framework provides a dynamic structure that can accommodate new developments while maintaining an unwavering commitment to patient welfare. This balanced approach ensures that the transformative potential of foundational models in healthcare can be realized while upholding the fundamental principles of medical ethics and human dignity.

\section*{Conflict of Interest Statement}
Dr. Bagci acknowledges the following COI: Ther-AI LLC. Dr. Wallace acknowledges the following COIs: Boston Scientific, ClearNote Health, Cosmo Pharmaceuticals, Endostart, Endiatix, Fujifilm, Medtronic, Surgical Automations, Ohelio Ltd, Venn Bioscience, Virgo Inc., Surgical Automation, and Microtek. All other authors declare no conflicts of interest. The funders had no role in the design of the study; in the collection, analyses, or interpretation of data; in the writing of the manuscript; or in the decision to publish the results.

\section*{Author Contributions}
DJ and UB conceived the presented idea, and GK, DY, AM, VC, EC, BA, MA, FM, RK, HS, DL, MB gave critical feedback about the study. NT, LP, SL, KB, EA, ZZ, BW, XZ, HP, AR, RH, MMR contributed to specific sections of the paper. UB took all the inputs from the authors prior to submission and finalized the manuscript. All authors discussed the varying aspects of ethical considerations in medicine and made sure there was a consensus. All authors contributed to the writing and revision of the paper, they all agreed with the final version.

\section*{Funding}
This project is supported by NIH funding: R01-CA246704, R01-CA240639, U01-CA268808, and R01-HL171376.

\bibliographystyle{unsrt}  
\bibliography{references}

\end{document}